# Unraveling the symmetry of $Al_5C_3N$

V. Shtender[1,*], C.S. Ong[2], P. Berastegui[1], O. Donzel-Gargand[3], J. Cedervall[1], C. Hervoches[4], P. Beran[4,5], O. Eriksson[2,6], and U. Jansson[1]

[1] Department of Chemistry, Ångström Laboratory, Uppsala University, S-75121 Uppsala, Sweden

[2] Department of Physics and Astronomy, Uppsala University, S-75120 Uppsala, Sweden;

[3] Division of Solar Cell Technology, Department of Materials Science and Engineering, Uppsala University, S-75121 Uppsala, Sweden

[4] Nuclear Physics Institute CAS, Rez 25068, Czech Republic

[5] European Spallation Source, ESS ERIC, S-221 00 Lund, Sweden

[6] WISE Wallenberg Initiative Materials Science, Uppsala University

* Corresponding author
vitalii.shtender@angstrom.uu.se

**Abstract**

The high-temperature ceramic compound $Al_5C_3N$ with promising application usage belongs to the scarcely studied Al–C–N system. It was originally reported as an ordered compound in the non-centrosymmetric space group *P*6$_3$*mc* and described as a nanolaminate with an –$Al_2C$–AlN–$Al_2C_2$– stacking sequence. The recently reported structural disorder in the related compound $Al_4SiC_4$ led us to question this proposed structure for $Al_5C_3N$ and investigate the possibility of a disordered structure in the centrosymmetric space group *P*6$_3$/*mmc*. In the present work, we employed different synthesis routes to maximize the yield and quality of the desired phase, and applied a variety of techniques to probe the $Al_5C_3N$ crystal structure. Our single-crystal X-ray diffraction analysis clearly indicates that the non-centrosymmetric space group *P*6$_3$*mc* must be rejected. From a joint refinement of single-crystal X-ray and powder neutron diffraction data, the occupancies of C and N were refined at two sites in *P*6$_3$/*mmc* resulting in the stacking sequence –$Al_2C$–Al(C/N)–$Al_2$(C/N)$_2$–. Furthermore, DFT calculations show that a centrosymmetric disordered structure described in a supercell has the lowest energy, 0.2 eV per formula unit, relative to the previously reported *P*6$_3$*mc* structure. The calculated band structure shows both direct and indirect band gaps which lead to implications for the physical properties. Finally, STEM analysis provides additional evidence that the crystal structure of $Al_5C_3N$ is better described in the centrosymmetric space group *P*6$_3$/*mmc*.

## 1. Introduction

Several ternary phases have been described in the Al-C-N system. The hexagonal phase $Al_5C_3N$ was first reported in 1935 by von Stackelberg et al.[1, 2], and in 1963, Jeffrey and Wu investigated the crystal structure of small crystals obtained from the walls of a carbon crucible after heating AlN powder to about 2000 °C in a nitrogen atmosphere.[3] They observed a series of phases with a chimney-like structure having the general formula $Al_{4+n}C_3N_n$ with $n$ = 1-4. The aluminum carbonitrides with an odd $n$ number, $Al_5C_3N$ and $Al_7C_3N_3$, were described with a hexagonal structure in space group $P6_3mc$ (#186), while the members with even $n$ number, $Al_6C_3N_2$ and $Al_8C_3N_4$, were described as rhombohedral in the space group $R\bar{3}m$ (#166). Based on the single crystal structure analyses of Jeffrey and Wu,[3, 4] the structure of $Al_5C_3N$ can be described as a nanolaminate with AlN layers between $Al_2C_2$ and $Al_2C$ blocks, whereas in $Al_8C_3N_4$, for example, the repeating sequence is $-Al_2C_2-$ AlN–AlN–$Al_2C$–AlN–AlN–. Thus, the $Al_{4+n}C_3N_n$ phases could be described as a series of compounds with the end members $Al_4C_3$ (for $n$= 0) and theoretically AlN (for $n$= ∞). The ordered structure of $Al_5C_3N$ can also be visualized with layers of polyhedra with C in tetrahedral or trigonal bipyramidal, and octahedral coordination, as in $Al_4C_3$, and N in tetrahedral coordination, as shown in Figure 1.

The stability of $Al_5C_3N$ has been confirmed experimentally[5-7] and according to a calculated phase diagram, this phase melts incongruently at 2253°C.[8] Samples with the $Al_5C_3N$ phase have been prepared from mixtures of $Al_4C_3$ and AlN by hot pressing in inert atmospheres[5, 6] or heating in a $H_2$ atmosphere,[7] by plasma jet cladding[9], and by heating $Al_4C_3$ in an $N_2$ atmosphere.[1] However, the existence of pure aluminum carbonitrides with $n > 1$ has been difficult to reproduce, and several authors have suggested that they are stabilized by impurities.[7, 8]

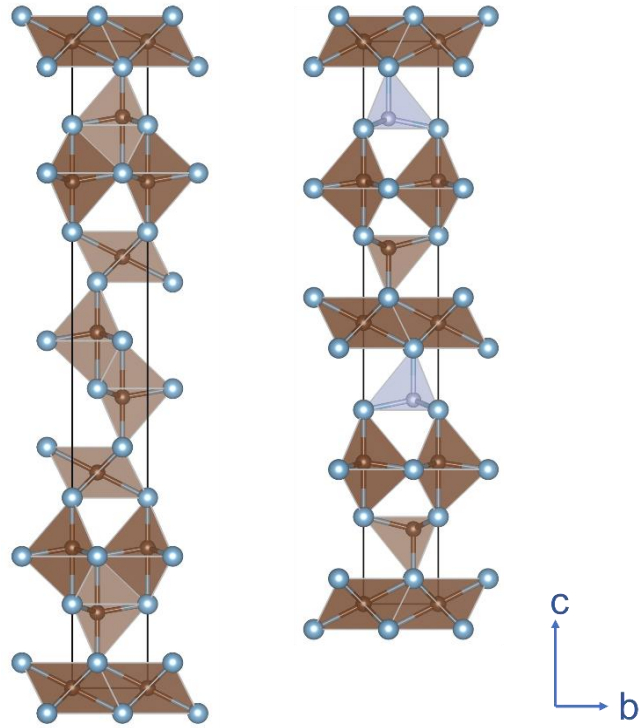

*Figure 1. Crystal structures of Al$_4$C$_3$[10] (left) showing the 5-fold coordination of C (brown) in –Al$_2$C$_2$– blocks between the –Al$_2$C– blocks with C in 6-fold coordination to Al (blue) atoms; Al$_5$C$_3$N (right) as reported by Jeffrey and Wu[3] with additional –AlN– layers (grey tetrahedra).*

Since other elements such as Si may play a role in stabilizing the Al$_{4+n}$C$_3$N$_n$ phases with $n > 1$, it is interesting to compare the Al–C–N system with the Al–C–Si system, where ternary phases with the general formula Al$_4$Si$_n$C$_{3+n}$ have been observed. First reported in 1961, the hexagonal unit cell of Al$_4$SiC$_4$ is very similar to that of Al$_5$C$_3$N.[6, 11] We have recently revisited the crystal structure of Al$_4$SiC$_4$ using a combination of experimental and theoretical methods and present a crystal structure with a mixed occupancy of Si and Al and the presence of Al vacancies, in contrast to the previously proposed ordered structure, which might have some interesting implications for the properties of Al$_4$SiC$_4$.[12]

Many applications have been proposed for phases in the Al–C–Si system, and it can be assumed that similar properties are also exhibited by the phases in the Al–C–N system. However, only one theoretical study of the electronic structure of Al$_5$C$_3$N has been published, indicating that this phase is a narrow band gap semiconductor.[13] Furthermore, a mainly covalent-ionic bonding with stronger Al–N bonds than Al–C bonds was predicted, consistent with a nanolaminated structure and possibly higher ductility than most MAX phases and Al$_4$SiC$_4$.[14]

A number of discrepancies in the structural determination of Al$_5$C$_3$N were reported and attributed to both experimental errors and stoichiometric defects.[3] Subsequent studies on this phase have used the original determination in the non-centrosymmetric space group *P6$_3$mc*. However, the same extinction conditions are observed in the centrosymmetric space group *P6$_3$/mmc* for a disordered structure, and the original structure proposed by Jeffrey and Wu[3] may be incorrect. The aim of this work is therefore to make a reassessment of this structure with a more detailed and complete study using both experimental and theoretical methods. We have used single-crystal X-ray diffraction, as well as neutron powder diffraction to study the structure of Al$_5$C$_3$N. Furthermore, we have used transmission electron microscopy (TEM) to gain additional information, and DFT calculations have been used to determine the most stable structural configuration of this phase.

## 2. Methods

Al$_5$C$_3$N was synthesized from Al$_4$C$_3$ (Alfa Aesar 99+ %) as powder packed in a cylindrical graphite crucible with 40 mm ID. The crucible was placed in the chamber of a graphite furnace from Thermal Technology LLC that was evacuated and refilled with Ar (6N) three times before N$_2$ (g) was mixed using mass flow controllers to obtain a 1% N$_2$/Ar atmosphere. After allowing the atmosphere inside the chamber to stabilize, the chamber was heated to 1950°C at a rate of 25°C/min and kept at that temperature for 60 min. The cooling rate was set at 75°C/min, but the furnace cooled naturally below approximately 1000°C. The partially sintered sample was covered with a thin layer of graphitic carbon that was scraped off before the rest of the sample was removed, ground, and packed in the graphite crucible. The heat treatment and this process were repeated five times, after which no further reaction of the remaining Al$_4$C$_3$ could be measured.

A small single crystal of Al$_5$C$_3$N was measured using a Bruker D8 single-crystal X-ray diffractometer with Mo Kα radiation (λ = 0.71073 Å). The diffractometer was equipped with an Incoatec Microfocus Source (IμS) and an APEX II CCD area detector. Single crystal X-ray diffraction (SCXRD) data reduction and numerical absorption corrections were performed using the APEX III software from Bruker.[15] Structure solution using the Superflip method, as well as the refinement, were carried out using JANA2020.[16]

Given the near identical atomic form factor of C and N for X-ray radiation (at s = 0, C – 6 e$^-$; N – 7 e$^-$), but significantly different for neutrons (scattering length, C – 6.64 fm; N – 9.36 fm), a neutron powder diffraction measurement was performed using the MEREDIT diffractometer at the Nuclear Physics Institute CAS in Rez, Czech Republic. A neutron beam with a wavelength of 1.46 Å was applied using a copper mosaic monochromator (reflection 220). A diffraction pattern in a 2θ-range of 4-144° with steps of 0.08° was collected at room temperature. The acquired powder diffraction patterns were analyzed with the software FullProf[17] using the Rietveld method.

Powder morphology and composition were studied using a ZEISS Leo 1550 field emission scanning electron microscope (SEM) equipped with an AZtec energy dispersive X-ray detector for spectroscopy analysis (EDS). Raman spectra were collected with a Renishaw inVia confocal Raman microscope using a 532 nm laser. IR spectra were collected using a Perkin Elmer Spectrum One instrument, equipped with a KBr/PE beam-splitter, DTGS/KBr detector and a Pike GladiATR diamond ATR unit, at a resolution of 2 cm$^{-1}$. The polycrystalline samples were also studied by powder X-ray diffraction (XRD) in θ-2θ mode using Cu Kα radiation. The measurements were done with a Bruker D8 Advance instrument equipped with a Lynxeye-XE detector and Ni filter. Quantitative phase analysis was done using the reference intensity ratio (RIR) method.[18] The results from these measurements are presented in the Supplementary Information.

The density functional theory (DFT) calculations were performed using Quantum ESPRESSO,[19] which uses a plane-wave basis set. The plane-wave cut-off for the DFT calculation was set to 85 Ry for the plane-wave expansion of the wave functions using the scalar-relativistic optimized norm-conserving Vanderbilt pseudopotential (ONCVPSP)[20] obtained from the PSEUDODOJO project.[21] The Perdew-Burke-Ernzerhof (PBE) functional within generalized gradient approximations (GGA) was used as the DFT exchange-correlation functional. For all structures, all components of all forces were minimized within the convergence threshold of $10^{-5}$ Ry per Bohr radius, and the total energy was also minimized within the convergence threshold of $10^{-8}$ Ry. Integrations over the reciprocal space were performed on 15×15×2 and 8×8×2 **k**-grids for the unit cell and (2×2)-supercell, respectively.

$Al_5C_3N$ crystals were also investigated by transmission electron microscopy (TEM). The lamellae were prepared using a Ga-based Focused Ion Beam (FIB) CrossBeam550 from Zeiss. The ion acceleration voltage was gradually reduced to 1 kV to minimize the polishing damage in the final lamella. A particular effort was made to include the c-axis of the crystal in the plane of the lamella. The TEM analyses were performed at 200 kV on a Titan Themis 200 from Thermofisher (Formely FEI) equipped with a Cs probe-corrector and a SuperX EDS system. The sample was loaded the day before the TEM study for improved stability. In Scanning (S)TEM, the high-angle annular dark field (HAADF) detector and the annular bright field (ABF) detector collected signals ranging between 70-200 mrad and 10-25 mrad, respectively. The simulated STEM images were calculated with a multi-slice approach using the software Dr. Probe[22] with the unit cell refined from neutron diffraction data as input.

## 3. Results and discussion

### 3.1 Synthesis of $Al_5C_3N$

$Al_4C_3$ is commonly used as a reactant in the preparation of MAX phases[23] and has been studied as a constituent in multicomponent systems and Al alloys.[24, 25] However, it is easily hydrolyzed[26] and can be detrimental to the mechanical properties if present as an impurity. $Al_4C_3$ decomposes at 2150°C into graphite and a liquid phase of C dissolved in Al[27], but a significant volatility of Al has been observed during annealing at temperatures above 1700°C.[28-30] The related $Al_5C_3N$ phase, on the other hand, is stable at ambient air and on cooling from high temperature, but the synthesis of this compound is not as straightforward. In the literature, two different routes have been used to synthesize $Al_5C_3N$: (i) heating of $Al_4C_3$ in an inert atmosphere with $N_2$ and (ii) a high-temperature reaction between $Al_4C_3$ and AlN. The first route was used by Stackelberg et al. in the 1930s by heating $Al_4C_3$ in an atmosphere of $N_2$ diluted in $H_2$ at temperatures below 2200°C [1]. $Al_5C_3N$ can be considered to be an intermediate phase in the nitridation of $Al_4C_3$ to AlN, and high temperatures or partial pressures of $N_2$ will increase the amount of AlN that forms. Their observations also suggested that the reaction started with a partial delamination of $Al_4C_3$ and sublimation before forming the carbonitride. The second route has been used with the sintering of mixtures of $Al_4C_3$ and AlN at atmospheric pressure[7] or by hot-pressing[6] at 1800°C for 30-60 minutes. However, the qualities of these samples are difficult to assess as no diffraction patterns were published. Reacting $Al_4C_3$ with AlN as a nitrogen source in inert atmospheres in route (ii) leads to the formation of the carbonitride at lower temperatures compared to route (i), and the reaction is apparently fast. However, $Al_5C_3N$ reacts with AlN to form $Al_6C_3N_2$, and impurities in the reactants or the atmosphere may result in the formation of an aluminum oxycarbonitride.[5, 31]

In our study, we have investigated both routes (i) and (ii) to determine the most efficient way to synthesize $Al_5C_3N$. Table 1 summarizes the results of phase analyses of $Al_4C_3$ samples that have been heated in different conditions using route (i), i.e., heating of $Al_4C_3$ in a $N_2$ atmosphere. All samples synthesized by this reaction are covered with a layer of graphitic carbon that can be mechanically removed, and the carbon content has thus not been considered. After heating in an inert Ar atmosphere (samples 1 and 2), no other phases are observed, but a weight loss of about 9% at 1900°C and 16% at 2000°C was measured due to the decomposition of $Al_4C_3$ and evaporation of Al. The $Al_5C_3N$ phase was observed after heating $Al_4C_3$ in atmospheres with $N_2$, and longer annealing times at lower $N_2$ partial pressures increased the fraction of the carbonitride phases (samples 3 to 8). Moreover, two AlN phases with the same crystal structure (wurtzite) but with different unit cell volumes, 41.75 and 42.19 Å$^3$, were found in all samples heated in an atmosphere with > 1.5% $N_2$ in Ar. The smaller cell is consistent with the reported unit cell for AlN and forms from the decomposition of $Al_5C_3N$ (sample 6), while the phase with the larger unit cell should form due to the nitridation of Al in nitrogen-rich atmospheres. The reason for this difference in unit cell volume is unclear. In summary, samples with $Al_5C_3N$ as the main phase (> 50 %) were obtained from the nitridation of $Al_4C_3$ in atmospheres with low partial pressures (1-1.5%) of $N_2$ and temperatures in the range 1950 - 2000°C. SEM images of the $Al_5C_3N$ crystals together with EDX maps, were collected (Figure SI1 and SI2, respectively).

*Table 1. Phase fractions determined using the RIR method from powder diffractograms in samples heated once at different temperatures and partial pressures of $N_2$.*

| Sample | Atmosphere | $Al_4C_3$:AlN | T (°C) | t (min) | Phase fractions (wt%) | | | | Reactions |
|---|---|---|---|---|---|---|---|---|---|
| | | | | | $Al_5C_3N$ | $Al_6C_3N_2$ | $Al_4C_3$ | AlN | |
| 1 | Ar | 1:0 | 1900 | 60 | | | 100 | | 1 |
| 2 | Ar | 1:0 | 2000 | 60 | | | 100 | | 1 |
| 3 | $N_2$ | 1:0 | 1900 | 60 | | | 30 | 70 | 1, 2 |
| 4 | $N_2$ | 1:0 | 2000 | 10 | 5 | | 90 | 5 | 1, 2, 4 |
| 5 | 10% $N_2$ | 1:0 | 1900 | 60 | 40 | 5 | 35 | 20 | 1, 2, 4, 6 |
| 6 | 4% $N_2$ | 1:0 | 2000 | 60 | 60 | | 20 | 20 | 1, 4, 5 |
| 7 | 1.5% $N_2$ | 1:0 | 2000 | 30 | 20 (70) | | 80 (30) | | 1, 4 |
| 8 | 1% $N_2$ | 1:0 | 2000 | 120 | 40 | | 60 | | 1, 4 |
| 9 | Ar | 2:1 | 2000 | 10* | 70 | | 10 | 20 | 1, 3 |
| 10 | Ar | 1:4 | 2000 | 60 | 20 | | | 80 | 1, 3 |
| 11 | $N_2$ | 1:1 | 2000 | 120 | 50 | 5 | | 45 | 1, 2, 3, 4, 6 |

The numbers in parentheses for sample 7 indicate the fractions obtained after a second sintering.
*Sample heated at about 65°C/min

(1) $Al_4C_3 \rightarrow 4\ Al + 3\ C$;  (2) $Al + ½\ N_2 \rightarrow AlN$;
(3) $Al_4C_3 + AlN \rightarrow Al_5C_3N$;  (4) $5/4\ Al_4C_3 + ½\ N_2\ (g) \rightarrow Al_5C_3N + 0.75\ C$;
(5) $Al_5C_3N + 2\ N_2\ (g) \rightarrow 5\ AlN + 3\ C$;  (6) $Al_5C_3N + AlN \rightarrow Al_6C_3N_2$

We have also investigated the formation of $Al_5C_3N$ using route (ii), i.e., heating of mixtures of $Al_4C_3$ and AlN in different ratios in a pure Ar or $N_2$ atmosphere (see samples 9-11 in Table 1). The largest fraction of $Al_5C_3N$ phase was obtained in a sample with a short dwelling time at high temperature in an Ar atmosphere which confirms the fast solid-state reaction between $Al_4C_3$ and AlN. A molar ratio of 2:1 was used to compensate for the partial decomposition of $Al_4C_3$ at high temperatures (sample 9). $Al_5C_3N$ is also formed from mixtures rich in AlN but in low yields (sample 10) which suggests that the graphitic carbon that forms from the decomposition of $Al_4C_3$ inhibits the solid-state reaction. The mixture heated in $N_2$ shows that all $Al_4C_3$ reacts after two hours and the observed AlN phase has formed from the nitridation of Al.

The reaction pathways to form $Al_5C_3N$ can thus be described as a fast gas-solid phase reaction involving the nitridation of $Al_4C_3$ or a solid-state reaction of $Al_4C_3$ with AlN depending on the reactants used. However, the latter reaction results in samples with both $Al_5C_3N$ and $Al_6C_3N_2$ phases due to the further reaction between $Al_5C_3N$ and AlN. In favor of purity, all subsequently experimentally characterizations discussed in this paper are performed on samples synthesized using route (i).

### 3.2 Single crystal X-ray diffraction of $Al_5C_3N$

Following route (i) above for the synthesis of a polycrystalline sample, we were able to isolate a few single crystals large enough for a structure determination using a single crystal diffractometer. As previously described, in 1963, Jeffrey and Wu suggested that $Al_5C_3N$ crystallized in the non-centrosymmetric $P6_3mc$ space group.[3] This structure, from a crystallographic point of view, is very similar to a structure in the centrosymmetric $P6_3/mmc$ space group. The difference between them is that due to the lack of inversion symmetry in $P6_3mc$, N atoms can fully occupy either of two four-fold sites resulting in an ordered arrangement, whereas in $P6_3/mmc$ these sites are equivalent and partially occupied by C and N atoms. A final assignment of the structure to one of them depends on the quality of the diffraction data, and Jeffrey together with Wu reported a large error in the $R$ factor, suggesting that the proposed $P6_3mc$ space group may be incorrect. To verify the space group symmetry, a small single crystal was selected and analyzed, and the space group test with the program JANA2020 suggested three equally valid space groups $P6_3/mmc$ (#194), $P\bar{6}2c$ (#190) and $P6_3mc$ (#186).

Table 2. Results of the crystal structure refinements of $Al_5C_3N$ in different space groups based on the SCXRD data.

| Space group (No.) | $P6_3mc$ (186) | $P\bar{6}2c$ (190) | $P6_3/mmc$ (194) |
|---|---|---|---|
| Al3 site occupancy at (1/3 2/3 z) | 1.0 at 2b | 0.5 at 4f | 0.5 at 4f |
| Anisotropic displacement $U_{33}$ for Al3 at 2b or 4f (Å$^2$) | 0.122(4) | 0.0119(18) | 0.0106(15) |
| Flack parameter | 0.2(5) | – | – |
| No. of independent reflections | 408 ($R_{eq}$ = 0.0462) | 296 ($R_{eq}$ = 0.0468) | 210 ($R_{eq}$ = 0.0471) |
| No. of reflections with $I > 2\sigma(I)$ | 327 ($R_{sigma}$ = 0.0234) | 240 ($R_{sigma}$ = 0.02) | 138 ($R_{sigma}$ = 0.0589) |
| Data/refined parameters | 408/27 | 296/16 | 210/17 |
| Goodness-of-fit on $F^2$ | 2.98 | 1.72 | 1.16 |
| Final R indices [$I > 2\sigma(I)$] | $R_1$ = 0.0548; $wR_2$ = 0.1417 | $R_1$ = 0.0312; $wR_2$ = 0.0868 | $R_1$ = 0.0281; $wR_2$ = 0.1013 |
| R indices (all data) | $R_1$ = 0.0692; $wR_2$ = 0.1456 | $R_1$ = 0.0428; $wR_2$ = 0.0900 | $R_1$ = 0.0444; $wR_2$ = 0.1047 |
| Largest diff. peak and hole (eÅ$^{-3}$) | 0.59 and -0.46 | 0.33 and -0.36 | 0.30 and -0.24 |

Results of the crystal structure refinements of $Al_5C_3N$ in $P6_3/mmc$ (194) have been deposited with the CCDC under deposition number 2504685.

The ordered structure model in $P6_3mc$ displays the expected layered structure with N in four-fold coordination as shown in Figure 1. However, the structure solution in this space group did not converge during refinement and resulted in an atomic displacement for Al at (0,0,¼), i.e. at the equatorial position of the trigonal bipyramid, about 10 times higher than the average value at other Al positions. Alternatively, in the space groups $P6_3/mmc$ and $P\bar{6}2c$, this Al site has to be described as a split site with half occupancy. In the refinement, N and C atoms were located at the same 4*f* Wyckoff position (⅓, ⅔, *z*) but residual electron density in the Fourier maps suggested that N also substituted for C at the 2*b* Wyckoff position (0,0,¼). A comparison of these possible models and well resolved peak intensities at high Q values, Figure 2, shows that the $P6_3mc$ model is not in agreement with the diffraction data from our crystal. Thus, the unstable refinement and large thermal displacement, see Table 2, suggest that the generally accepted space group $P6_3mc$ must be rejected. Moreover, the positive Flack parameter of + 0.2 for the $P6_3mc$ structure model indicates a possible inversion twin component. Subsequently a joint refinement was done using SCXRD and powder neutron data in the centrosymmetric space group, $P6_3/mmc$, see below, and parameters from the final refinements can be found in Table 2.

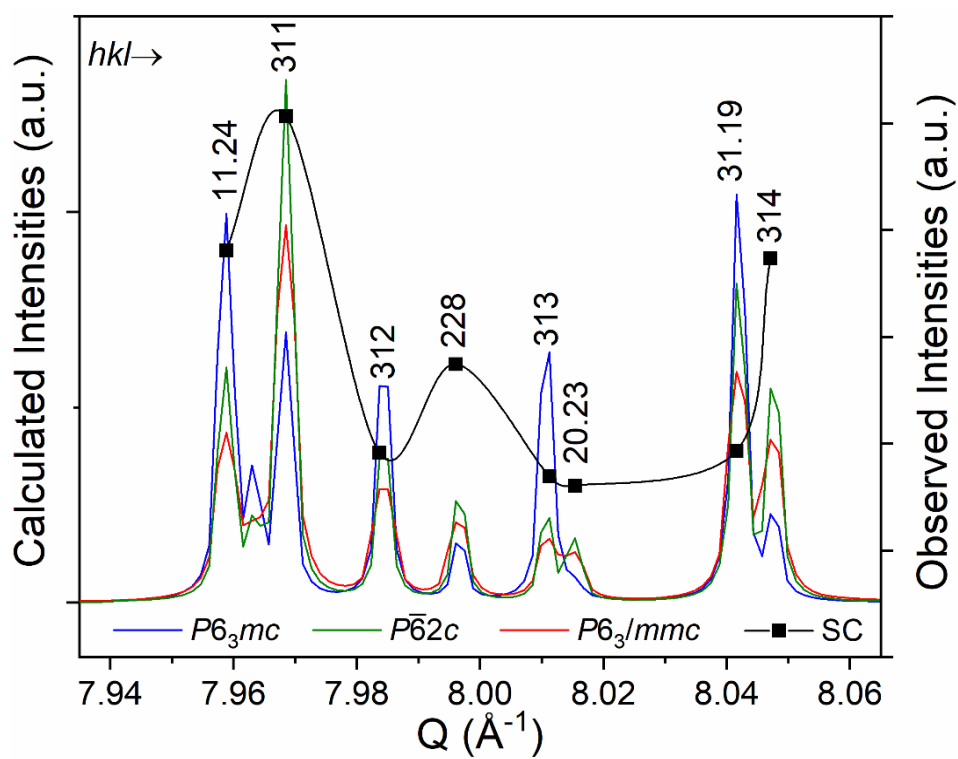

*Figure 2. Comparison of the observed single crystal peak intensities with the calculated intensities from the three structural models in space groups $P6_3mc$ (#186), $P\bar{6}2c$ (#190) and $P6_3/mmc$ (#194).*

### 3.3 Powder neutron diffraction of Al$_5$C$_3$N

The difference in neutron scattering length for C and N allows for the determination of their occupancies with less ambiguity. As observed during the single crystal structure solution, in the preliminary refinements of the structural model of Al$_5$C$_3$N in the *P6$_3$mc* space group, the position and displacement parameter for Al at *z* ≈ 0.25 were unstable and the disordered structure model in *P6$_3$/mmc* was tested. Rietveld analyses resulted in similar models with N substituting partially at one or two C sites and in order to determine the best solution, a joint refinement with SCXRD data was done. The final model takes into account the residual electron density that was found in the Fourier maps, the split Al site and shows that N partially occupies two sites with a restraint to the nominal stoichiometry. The unit cell parameters are *a*= 3.2829(5), *c*= 21.604(5) Å and *V*= 201.64(6) Å$^3$, and the final fit from the Rietveld refinement of the powder diffraction data is shown in Figure 3. The structural model of Al$_5$C$_3$N in the *P6$_3$/mmc* space group based on the joint refinement of the SCXRD and neutron diffraction data, Table 3, is shown in Figure 4 and calculated distances are shown in Figure SI3.

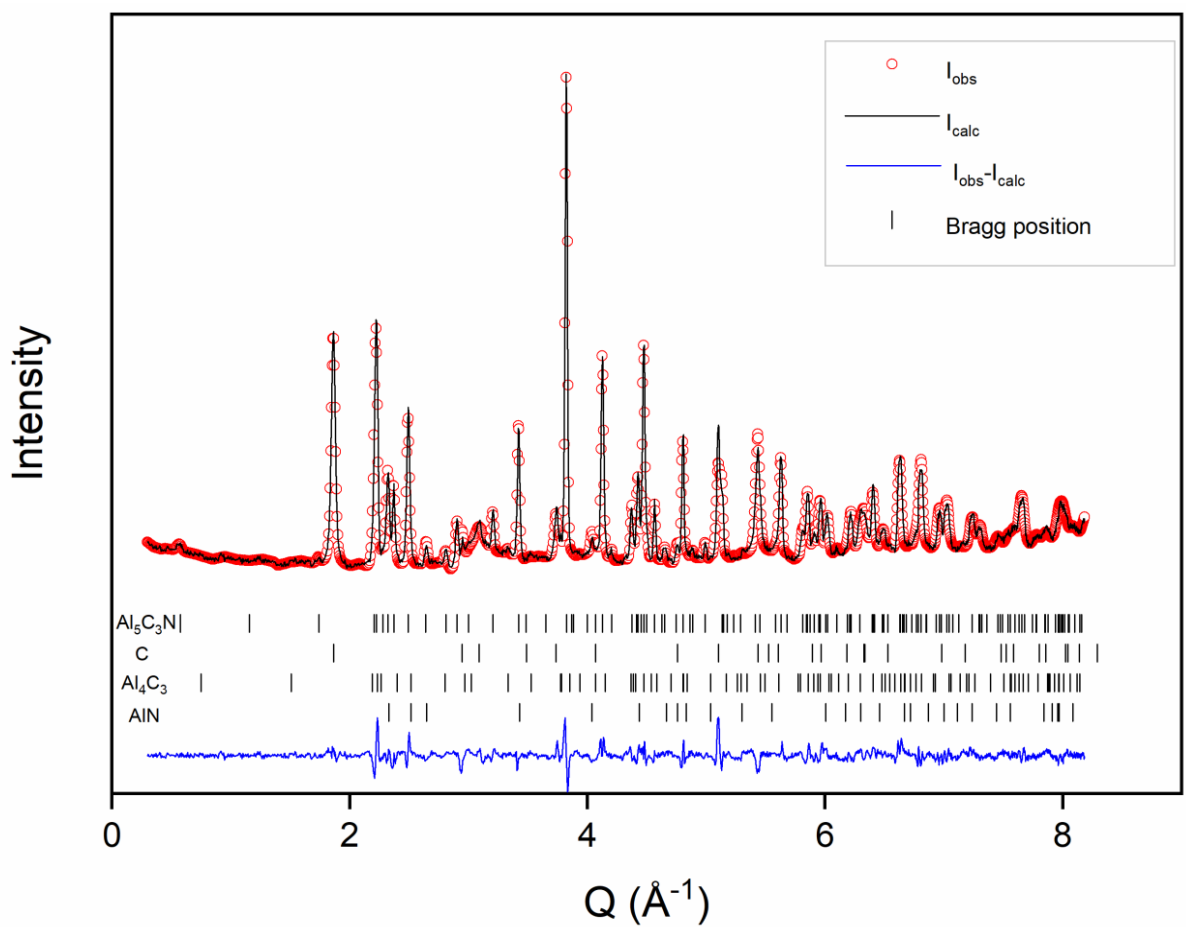

*Figure 3. Results from Rietveld analysis of room temperature powder neutron diffraction data (λ= 1.46 Å) for Al$_5$C$_3$N in P6$_3$/mmc. Total $\chi^2$ = 6.9%, R$_{wp}$= 4.5%, Al$_5$C$_3$N phase R$_{Bragg}$ = 5.5%. Phase fractions: Al$_5$C$_3$N 69.8%, C 24.5 %, Al$_4$C$_3$ 3.5%, AlN 2.2%.*

*Table 3. Refined atomic positions and temperature factors from joint refinement of the SCXRD and powder neutron diffraction data for $Al_5C_3N$ in space group $P6_3/mmc$ (#194).*

| Site | Wyckoff | x | y | z | $U_{iso}$ (Å$^2$) | Occupancy |
|---|---|---|---|---|---|---|
| Al1 | 4f | 1/3 | 2/3 | 0.04537(7) | 0.0079(8) | 1 |
| Al2 | 4e | 0 | 0 | 0.15384(8) | 0.0091(8) | 1 |
| Al3 | 4f | 1/3 | 2/3 | 0.2373(1) | 0.0049(8) | 0.5 |
| C1 | 2a | 0 | 0 | 0 | 0.0064(6) | 1 |
| M2 | 4f | 1/3 | 2/3 | 0.1329(2) | 0.0087(6) | 0.61C+0.39N |
| M3 | 2b | 0 | 0 | 1/4 | 0.0061(6) | 0.77C+0.23N |

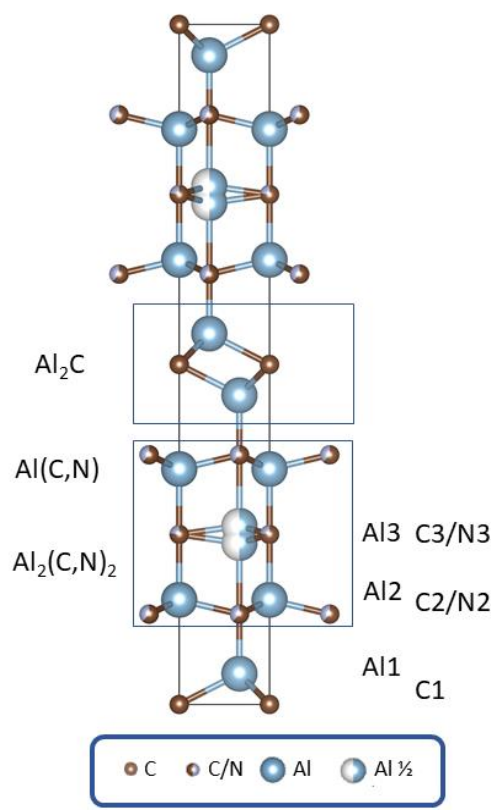

*Figure 4. Structure model of $Al_5C_3N$ in the $P6_3/mmc$ space group based on our data. The figure also shows the alternating slabs of $Al_2C$-$(AlN$-$Al_2C_2)$- that in this space group are disordered due to partial occupancies.*

### 3.4 Possible structural models of Al₅C₃N

Based on the results from the single crystal X-ray and powder neutron diffraction studies, the structure is better described as a disordered structure in the centrosymmetric *P6₃/mmc* space group. This disorder has been modelled in similar compounds with inversion twins, i.e. with the combination of two non-centrosymmetric structures related by an inversion center in an alternative structure solution based on an inversion twin. This type of twinning would not be evident from diffraction data as there are no significant differences in the calculated peak intensities between the structural models. However, an inversion center can be confirmed by measurements of physicochemical properties[32] or measurements of the active modes in Raman and IR spectra, see Figure SI5. Austerman et al., have shown that such twins can be created in wurtzite-type crystals during crystal growth with impurities or crystal strains, e.g. oxygen in AlN or strain gradients in BeO.[33] A similar solution has been proposed for the structure of $Al_3BC$ from X-ray diffraction data,[34] with a centrosymmetric model and the equatorial Al position of the trigonal bipyramid split along the *c*-axis. The origin of the static disorder was attributed to twinning and a later computational study proposed a structure modelling this disorder.[35] In these cases, models with split positions are the average structure as observed in a polycrystalline material but the split site can also be described using a propagation vector in a lower symmetry orthorhombic space group that simulates the movement of this atom.

A possible pathway to the formation of such inversion twins is outlined in Figure 5. $Al_4C_3$ can be described as a layered structure consisting of alternating $Al_2C$ and $Al_2C_2$ units. It is possible to form the $Al_5C_3N$ phase by inserting AlN layers between these units in two ways as illustrated in Figure 5. The stacking sequence can then be either $-Al_2C-AlN-Al_2C_2-$ (denoted α-$Al_5C_3N$) or $-Al_2C_2-AlN-Al_2C-$ (denoted β-$Al_5C_3N$). The α-$Al_5C_3N$ sequence is, in fact, the non-centrosymmetric $P6_3mc$ crystal structure of Jeffrey and Wu[3] and even though α-$Al_5C_3N$ and β-$Al_5C_3N$ are separately non-centrosymmetric, they are related by inversion symmetry. The combination of both stacking sequences will produce a pseudo-symmetric structure which would be identified as $P6_3/mmc$ in our X-ray and neutron diffraction studies above. In each case, an $Al_2C$-$Al_2C_2$ unit separate the AlN layers, keeping two AlN layers as far away from each other as possible and ensuring that the AlN layers are maximally dispersed.

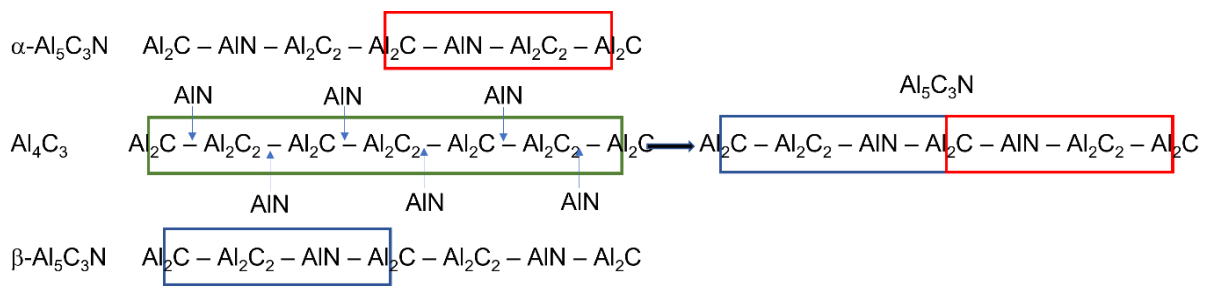

*Figure 5. Models for the formation of inversion twins in the $Al_5C_3N$ structure formed from the insertion of AlN in $Al_4C_3$ (left) and the $Al_5C_3N$ twin formed by a combination of α-$Al_5C_3N$ and β-$Al_5C_3N$ (right).*

However, a potential problem with the twinned structure outlined in Figure 5 is the formation of a twin boundary at an energy cost. A second possibility is that both α-$Al_5C_3N$ and β-$Al_5C_3N$ can coexist as a locally disordered phase within a supercell. This disorder generally makes the structure energetically less favorable unless the disorder gives rise to a periodic lattice distortion that extends beyond the unit cell, akin to a charge-density wave, which will lower the energy of the overall structure and induce a spontaneous symmetry breaking. In the next section, the stability of these two possibilities is investigated using DFT.

### 3.5 Theoretical calculations

In order to correctly describe the structure of the $Al_5C_3N$ phase and to understand the origin of the inversion symmetry, we begin by calculating the DFT energy of the $P6_3mc$ structure[3] (denoted as "Literature" in Figure 6d). First, we explore the possibility that the inversion symmetry that was determined from the diffraction experiments arises from naturally occurring inversion twins at the energetic cost of forming twin boundaries. We investigated all possible periodic positions of the twin boundary along the $c$-axis and found that when the twin boundary is located as shown in Figure 6a, each boundary would have the lowest formation energy of 0.6 eV per unit cell in the $ab$-plane, which is still prohibitively large and unlikely to be formed experimentally. In this work, the formation energy is defined as the DFT-calculated total energy of the configuration minus the DFT energy of the structure reported in the literature (see Fig. 6d).

We then calculate the formation energy needed to reorder the C/N-planes within the unit cell, with respect to the $P6_3mc$ structure. All permutations are considered and we will first discuss the two configurations (Figure 6b, Figure 6c) where each AlN-plane is formed along the surface of the $Al_2C$ slabs. In the first configuration (Figure 6b), we restore the inversion symmetry by ordering the AlN-planes such that they become related by inversion symmetry. In this configuration, the two N-planes are closer to each other than in the $P6_3mc$ structure. This configuration has a positive formation energy per $Al_5C_3N$ formula unit (*f.u.*) of 0.1 eV, suggesting that it is energetically more favorable to keep the AlN-planes apart. We hypothesize that this is because N has a lower electron affinity than C, and is unable to accept all the electrons that Al would like to donate in order to maintain charge neutrality. This would lead to the *n*-type doping of Al and the *p*-type doping of C. The lack of a complete shell leads to higher energy because it lacks the exchange-correlation stabilization and minimized repulsion of a closed-shell configuration. The system is thus more reactive and becomes less stable.

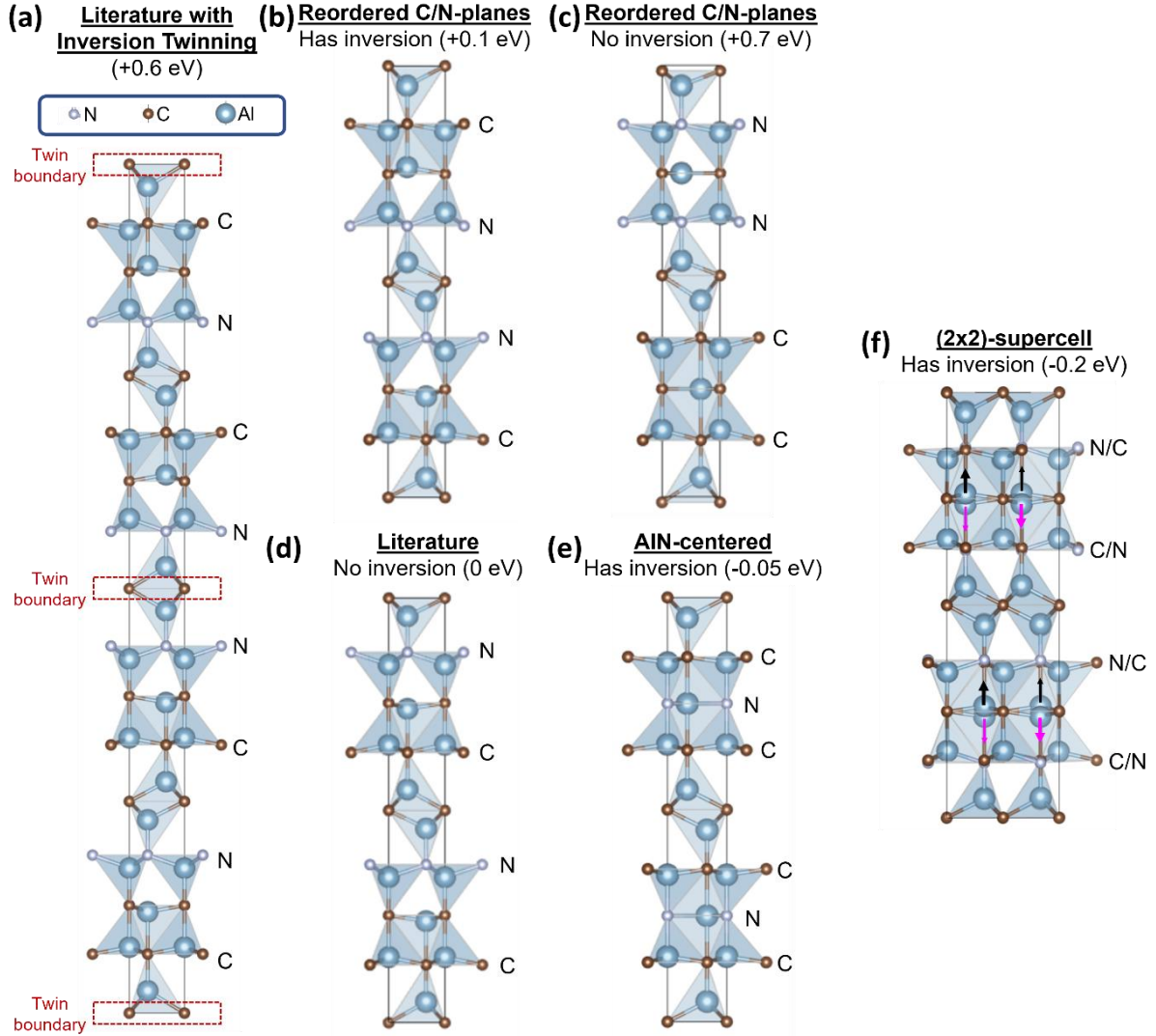

*Figure 6. Candidate structures of Al$_5$C$_3$N and their formation energies per f.u. relative to the published P6$_3$mc structure of Jeffrey and Wu[3] as shown in (d), which contains two f.u. (a) shows the published structure twinned along the c-axis and its energetically unfavorable twin boundaries. (b) and (c) show the published structure with two other possible site occupancies for N related by inversion symmetry in (b) and or not in (c). Both (b, c) are energetically less favorable than the published structure in (d), which has its AlN-planes maximally separated. (e) shows the Jeffrey and Wu[3] structure with an AlN-plane centered in the middle of the slab, (f) shows our proposed structure, which restores the inversion symmetry via the even atomic distributions of C/N within the same plane. This structure has the lowest energy due to spontaneous symmetry breaking within the (2×2)-supercell.*

To verify our hypothesis, we calculated the band structure of this configuration in Figure 7a and compared it against that of the *P6$_3$mc* structure in Figure 6b. Indeed, we found states at the Fermi level due to *p* and *n*-types doping for the more unstable structure. We further test our hypothesis by reordering the AlN-planes in a second configuration (Figure 6c), such that both the AlN-planes are now even closer. Not only does this configuration not have an inversion center, the formation energy is even larger than the first configuration (0.7 eV per f.u.), and the *p-/n*-types doping becomes more severe (Figure 7b). Having established that the AlN-planes prefer to be located as far away from each other as possible, we question why the AlN-planes would preferentially locate along one of the two surfaces of the -$Al_2C$-$Al_2C_2$- slabs, as suggested by Jeffrey and Wu. If each AlN-plane is located in the middle of the slab, not only would the overall structure have a higher symmetry belonging to the space group $P6_3/mmc$, this structure will also have an inversion symmetry and the space group observed in the X-ray diffraction experiments. In fact, our calculations show that such a structure would even have a negative formation energy of -0.05 eV/f.u. (Figure 6e) compared to the reference structure (Figure 6d), and is, therefore, more stable than the $P6_3mc$ structure proposed by Jeffrey and Wu.[3, 4] Nonetheless, our diffraction experiments also confirm that this hypothetical ordered structure is not a desirable solution.

Finally, we consider the possibility in which lattice modulation extends the periodicity beyond that of the unit cell along the *ab*-plane. This can induce spontaneous symmetry breaking that lowers the energy of the overall structure, akin to a charge-density wave. To this end, we created a (2×2)-supercell with even atomic distributions of C/N within the same plane (Figure 6f) and of all the structures we have evaluated, this structure has the lowest energy, that is 0.2 eV/f.u. lower than the published structure.[3] The formation energy of the structure is not only negative, confirming its thermodynamic favorability, but is also large in magnitude. In Figure 6f, we see that the checkered atomic distribution of C and N creates modulations in the local chemical environment that cause Al in the now $Al_2CN$ slab to be displaced slightly upward (black arrows) and downward (magenta arrows) in the *c* direction. Such atomic displacements are confirmed experimentally in our diffraction studies as a split position, and were also observed for $Al_4SiC_4$,[12] for which we also reassessed its crystal symmetry. Summarizing, inversion symmetry can be restored via the fractional occupation of Al(C,N) layers (Figure 6f). Such a structure can be interpreted as a disordered superposition of the (1×1)-unit cells that individually do not have inversion symmetry, as shown in Figure 6d. Indeed, the experimental results indicate additional disorder with N substituting at the two sites indicated in Fig. 6e and 6f, both energetically more favorable than the literature structure in Fig. 6d. However, due to computational constrains, the modelling of a structure with this degree of disorder was not carried out.

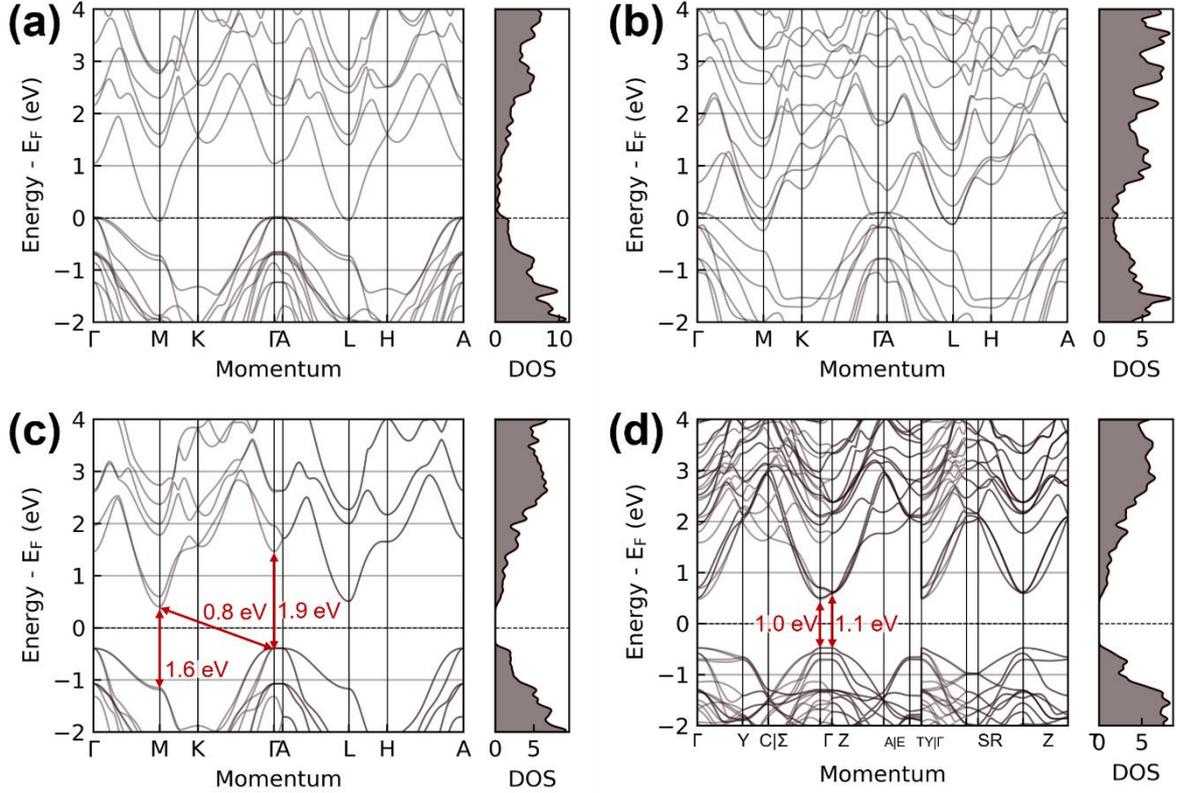

*Figure 7 (a-d) Band structures of the structures shown in Figure 6(b,c,d,f), respectively. Each band structure is accompanied by the density of states (DOS) on the right that is normalized to the number of atoms within the unit cell of $Al_5C_3N$.*

In Figure 7 we show the calculated electronic structure of the structures considered for $Al_5C_3N$ compound. Note that depending on crystal structure, the electronic structure is semi-conducting, semi-metallic or metallic. Figure 7c shows also the calculated DFT bandgap and we see that the *P6₃mc* structure has a direct gap of (1.6-1.9 eV) and an indirect gap of 0.8 eV, in agreement with Ref.[13] In the 2×2 supercell, Figure 7d, the conduction bands were folded into the Γ point, and the breaking of discrete translational symmetry has allowed for direct transition of 1.0 eV at the Γ point. Nonetheless, we note that since DFT bandgaps are not quasiparticle bandgaps, they underestimate the experimental bandgap of 2.2 eV, as expected from calculations of DFT level.[36]

## 3.6 Scanning transmission electron microscopy (STEM)

Our experimental results as well as the DFT calculations in section 3.5 clearly suggest that the generally accepted crystal structure for $Al_5C_3N$ with an ordered and non-centrosymmetric $P6_3mc$ (#186) space group is not correct, and that a centrosymmetric and disordered structure in $P6_3/mmc$ (#194) is a better fit to our results. To confirm this conclusion, a STEM study was carried out on crystals of $Al_5C_3N$. Experimental and simulated HAADF STEM images are shown in Figure 8.

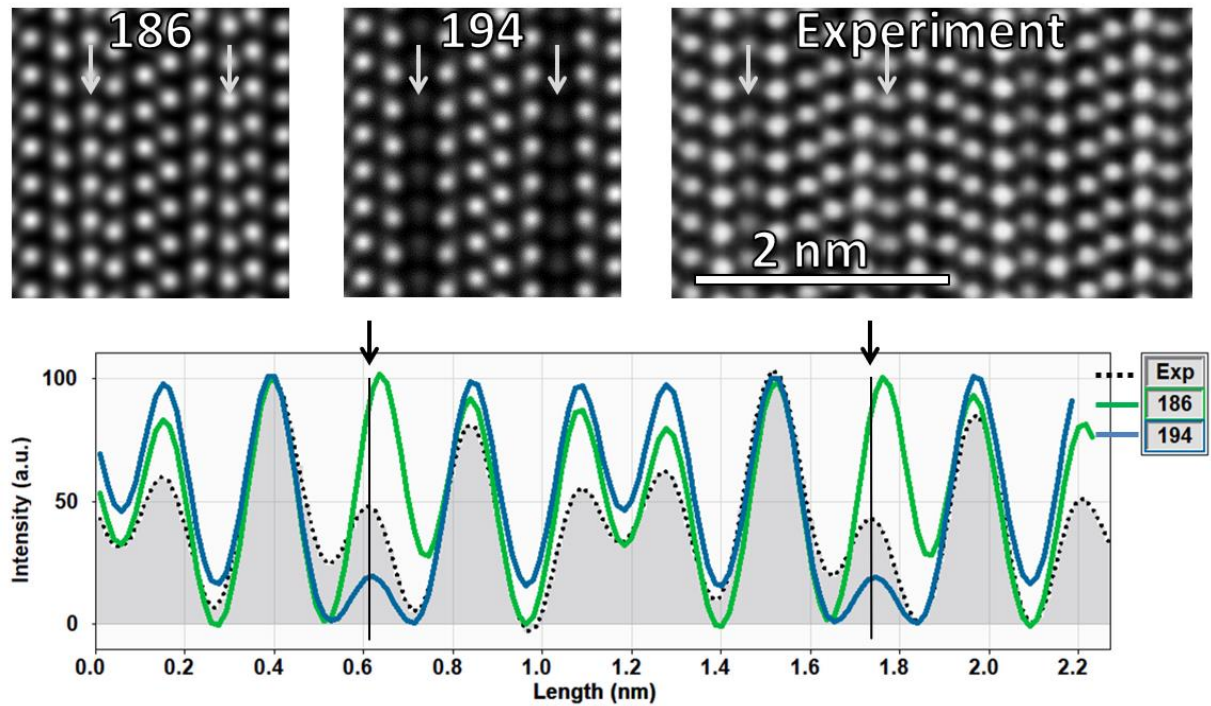

*Figure 8. (top) simulated HAADF STEM images from the model with space groups $P6_3mc$ (#186) and $P6_3/mmc$ (#194) using Dr.Probe software and a frozen lattice procedure (50 frozen states and total thickness of 20 nm) compared to the experimental one. (Bottom) Compilation of the horizontal line profiles of the HAADF intensities integrated over the height of the image. The vertical arrows in the intensity profiles and in the HAADF images point to the same double-Al atomic plane. For an easier comparison, intensity curves were background corrected and normalized at the Al(C,N) atomic plane (close to 0.4 nm).*

While the models $P6_3mc$ (#186) and $P6_3/mmc$ (#194) are closely related, some of the lattice plane spacings are slightly different due to the splitting of the Al site in $P6_3/mmc$. Deviations for model $P6_3mc$ are evident in Figure 8 (bottom) as indicated by the vertical lines at about 0.6 and 1.75 nm corresponding to the split Al sites, where $P6_3/mmc$ is closer to the experiment. Concerning the change in HAADF intensities, one can see that none of the current lattices perfectly fit the experiment, but once again the model in $P6_3/mmc$ is closer, with intensities being higher at the Al(C,N) planes and lower at the split Al ones. The intensity at the central $Al_2C$ plane also decreases for our model however not enough to match the experimental results. In other words, the apparent density of these lattice planes do not reflect exactly the STEM experiment and could indicate a different amount of vacancy as discussed in a previous work for $Al_4SiC_4$.[12]

The analysis of the chemical composition using EDS is presented Figure 9. The HAADF profile shows a similar sequence to the HRSTEM measurements in Figure 8 despite of the lower resolution due to the decreased sampling rate and larger electron probe utilised to get a useful count rate for the chemical mapping. In spite of the low X-ray counts, both integrated profiles for C and N feature clear repetitive sequences that can be correlated to the atomic model $P6_3/mmc$ (#194). Indeed, N is localized at Al(C,N) planes and C is localized at Al$_2$C planes. The fact that the quantified amounts of C nor N do not fall to zero even where it would be expected from the model is not surprising and can be explained by an insufficient spatial resolution. First, the higher beam current used for the measurement (~230 pA) translates as a larger probe size which enhances the EDS signal but degrades the spatial resolution as the tail of the gaussian probe would always slightly excite neighbouring atomic planes. Second, binning of the recorded data was used to enhance the signal-to-noise ratio which also broadens the spatial resolution. The thickness of the lamella has also been reported to rapidly reduce the chemical composition spatial resolution by broadening signals from one atomic plane to neighbouring ones. As a consequence, the peak maxima (and valley minima) are weighted by neighbouring pixels and lose amplitude, and the detectability criterion as described by Lu et al.[37] may not be fulfilled for the shorter interatomic distances, insufficient to resolve each planes properly. However, by accepting a lower signal to noise ratio and decreasing the binning level, the C signal start to emerge at several Al split sites as presented in Figure SI6, supporting further the refined model presented in this work.

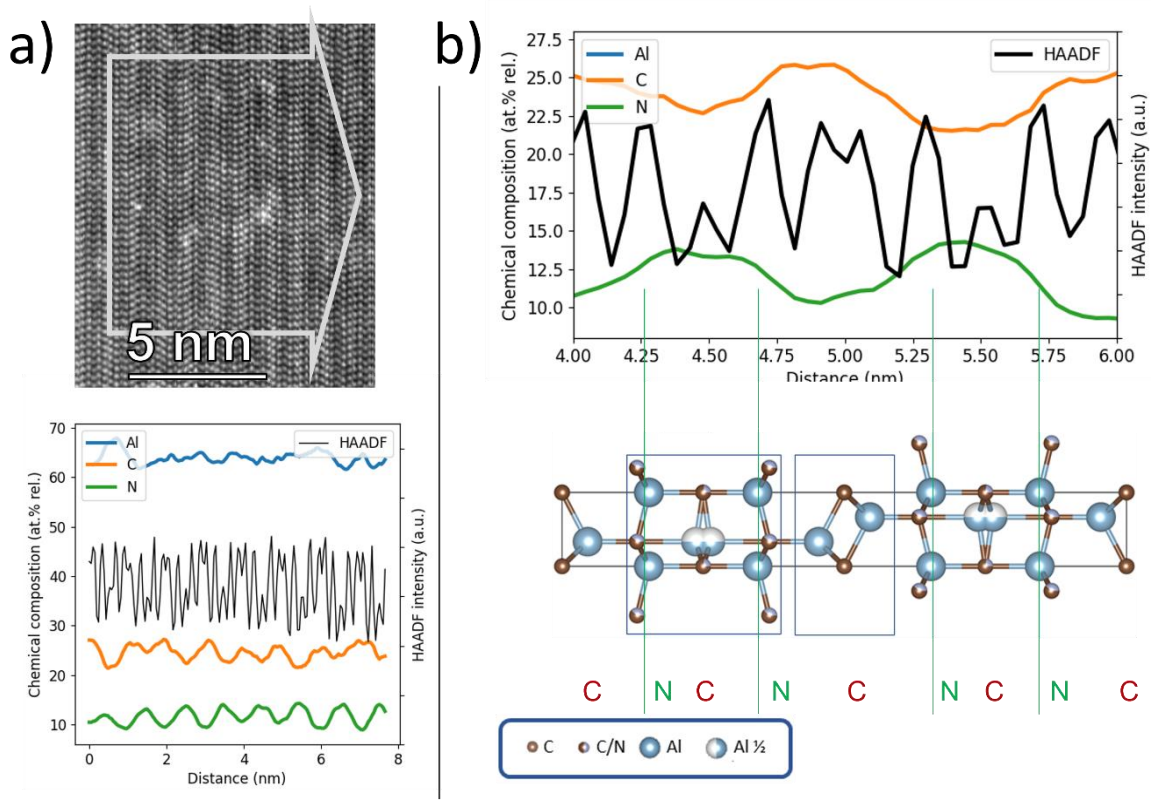

*Figure 9. (a) top, STEM HAADF survey image of the EDS mapped region, bottom, the corresponding integrated profiles of the chemical composition showing repetitive patterns for Al and N signals (b) top, cropped integrated profiles to compare it to our atomic model as presented in Figure 4. EDS measured in one single scan; the signal is integrated over a width of 10 nm as indicated by the grey arrow on the HAADF survey image.*

## 4. Conclusions

We have thoroughly revisited the crystal structure of the $Al_5C_3N$ compound. Based on the variety of experimental and theoretical techniques employed in this work, we conclude that $Al_5C_3N$ samples can be prepared from $Al_4C_3$ at high temperatures and $N_2$ partial pressures close to 1%. $Al_5C_3N$ crystallizes in a centrosymmetric space group (*P6₃/mmc* (#194)) rather than the previously reported non-centrosymmetric one (*P6₃mc* (#186)) with the following observations supporting this conclusion:

- SCXRD shows that the variation of observed peak intensities follows the same trend as expected for the centrosymmetric space group, while refinements in the non-centrosymmetric space group do not converge.
- Joint refinement of SCXRD and neutron powder diffraction data indicates that N and C are intermixing at the 4*f* and 2*b* Wyckoff sites.
- The possibility of inversion twinning, which could produce a pseudo-symmetric structure, would be energetically unfavorable.

- DFT calculations also demonstrate that a centrosymmetric structure has the lowest formation energy but it was not feasible to perform calculations on a structure with N disordered over two sites.
- STEM results reveal a variation in intensities consistent with SCXRD, confirming that the crystal structure of $Al_5C_3N$ is disordered. STEM HAADF combined with EDS further shows C and N compositional variation, in agreement with the diffraction results.

Our findings are expected to support future investigations of layered structures in the Al–C–N and Al–C–Si–N systems. The revised disordered crystal structure of the $Al_5C_3N$ compound will be crucial for theoretical predictions of its physical properties.

**Acknowledgements**

The Swedish Research Council (VR) is acknowledged for financial support (Grant 2022-03120). MyFab Uppsala is acknowledged for access and experimental support to the electron microscopy facilities. MyFab is funded by the Swedish Research Council (2020-00207) as a national research infrastructure. The authors acknowledge the financial support from the MEYS CR (Project OP JAK FerrMion, No. CZ.02.01.01/00/22_008/0004591). Neutron diffraction measurements were carried out at the CANAM infrastructure of the NPI CAS Rez, using the CICRR infrastructure supported by MEYS project LM2023041. O.E. acknowledges support from the Swedish Research Council, The Knut and Alice Wallenberg Foundation, eSSENCE, StandUp, the ERC (synergy grant FASTCORR, project 854843) and WISE-Wallenberg Initiative Materials Science, funded by the Knut and Alice Wallenberg Foundation.

**Supplementary information**

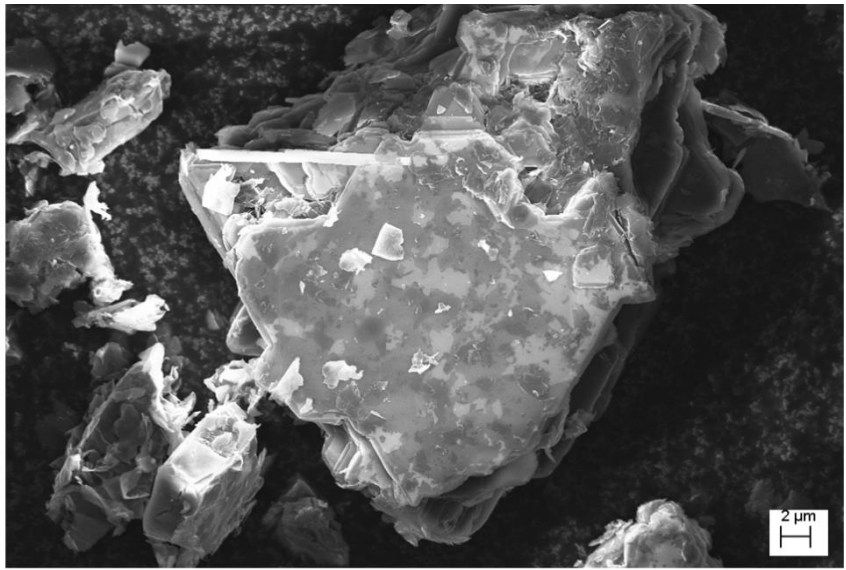

*Figure SI1.* SEM image of an Al$_5$C$_3$N crystal.

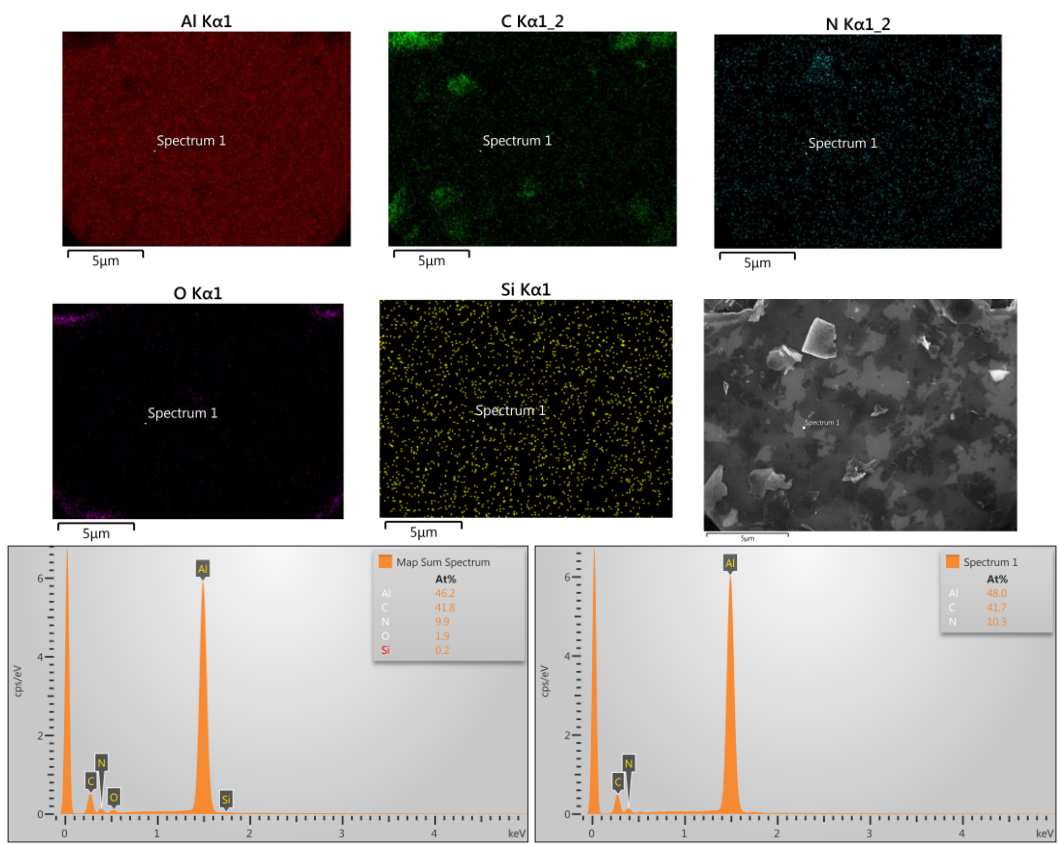

*Figure SI2.* EDS analysis of an Al$_5$C$_3$N crystal. The measured oxygen content is very low and C impurities and an AlN flake are visible.

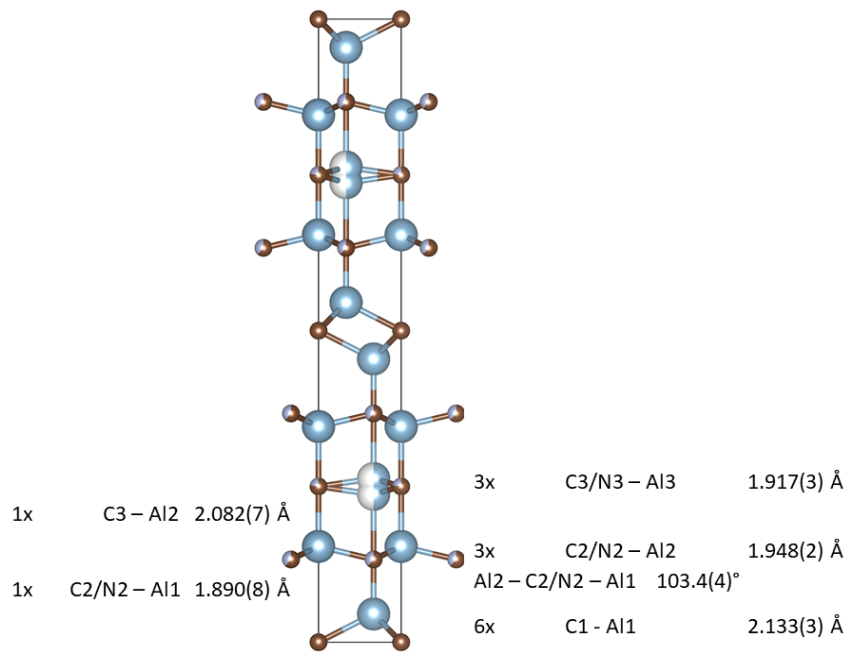

| | | | |
|---|---|---|---|
| 1x | C3 – Al2  2.082(7) Å | 3x  C3/N3 – Al3 | 1.917(3) Å |
| 1x | C2/N2 – Al1  1.890(8) Å | 3x  C2/N2 – Al2 | 1.948(2) Å |
| | | Al2 – C2/N2 – Al1 | 103.4(4)° |
| | | 6x  C1 - Al1 | 2.133(3) Å |

**Figure SI3.** Bond distances in Å and tetrahedral angles calculated for the $Al_5C_3N$ structure model in $P6_3/mmc$ from neutron diffraction data. The bonding is illustrated for C sites (brown) and C/N sites (blue/brown) while Al sites (blue) are shown with larger spheres.

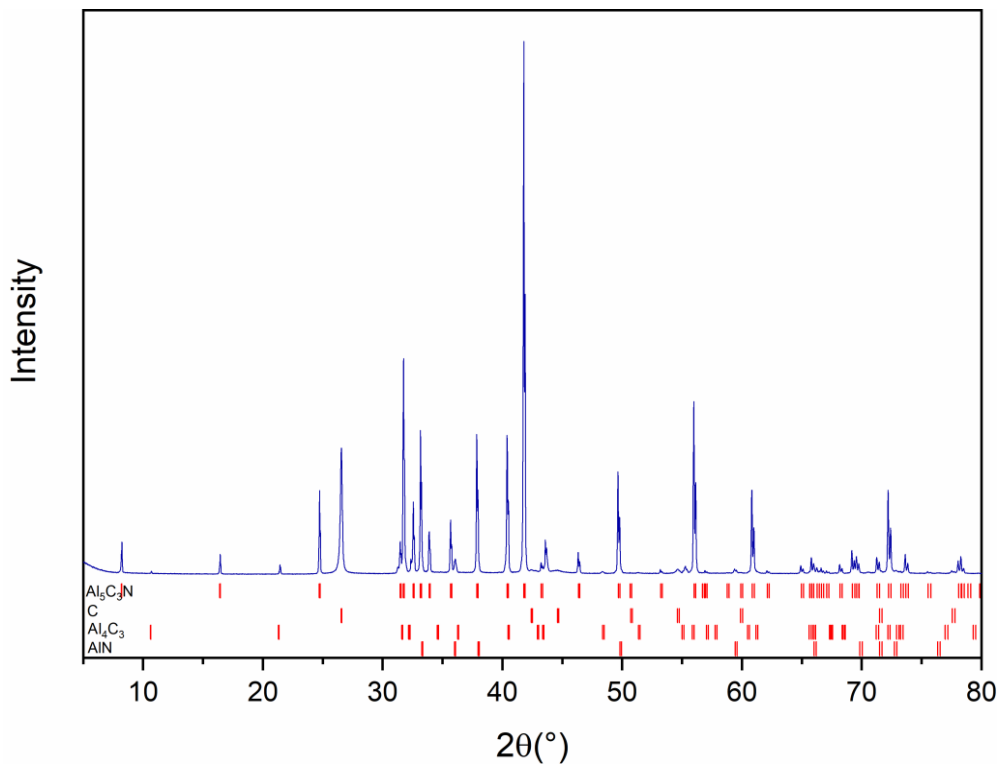

**Figure SI4.** X-ray diffractogram of the $Al_5C_3N$ sample with 2H-C as a second phase (peak at 26.6°) and $Al_4C_3$ and AlN as impurities. The main phase is $Al_5C_3N$ with a= 3.2833(3) and c= 21.618(3) Å.

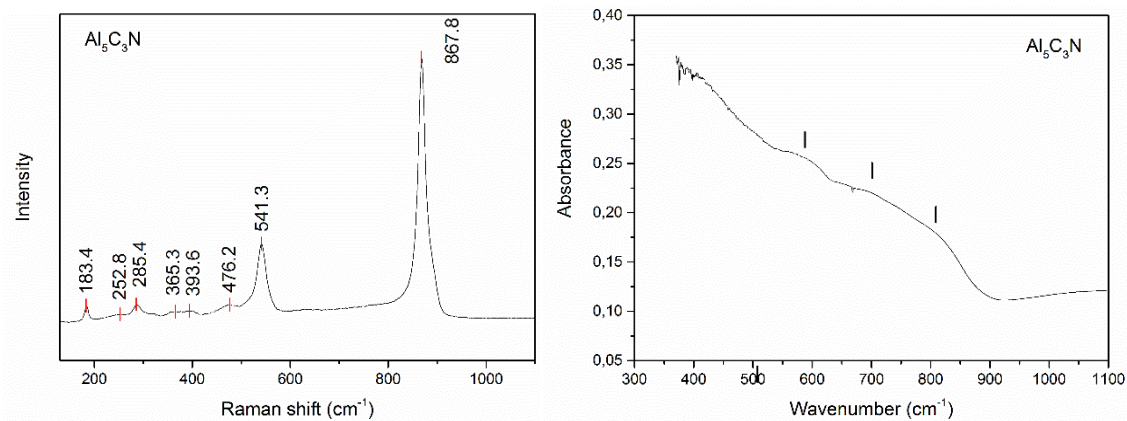

*Figure SI5.* Raman and ATR (attenuated total reflection) spectra from a polycrystalline $Al_5C_3N$ sample. The Raman spectrum of $Al_5C_3N$ shows strong A1 and E1 modes. The peak at 365 cm$^{-1}$ could be assigned to a silent B2 mode and has been attributed to impurities.[SI1] The IR spectrum shows weak overlapping modes and the two strong A1 modes observed in the Raman spectrum (at 541 and 868 cm$^{-1}$) are not observed in the IR spectrum (520-630 and 775-875 cm$^{-1}$) which would indicate a centrosymmetric structure.

*Table SI1.* Experimental positions in cm$^{-1}$ of $Al_5C_3N$ Raman modes and IR (ATR: Attenuated Total Reflection, marked positions in the figure) modes.

| $\omega_{exp}$ Raman | $\omega_{exp}$ ATR | Symmetry |
|---|---|---|
| 183 | | E1 |
| 252 | | E2 |
| 285 | | E1 |
| 365 | | |
| 393 | | A1 |
| 476 | | E1 |
| 541 | 590 | A1 |
| | 740 | E1 |
| 868 | 810 | A1 |

SI1    L. Pedesseau, O. Chaix-Pluchery, M. Modreanu, *et al.*. *J Raman Spectrosc* 2017, **48**, 891-896.

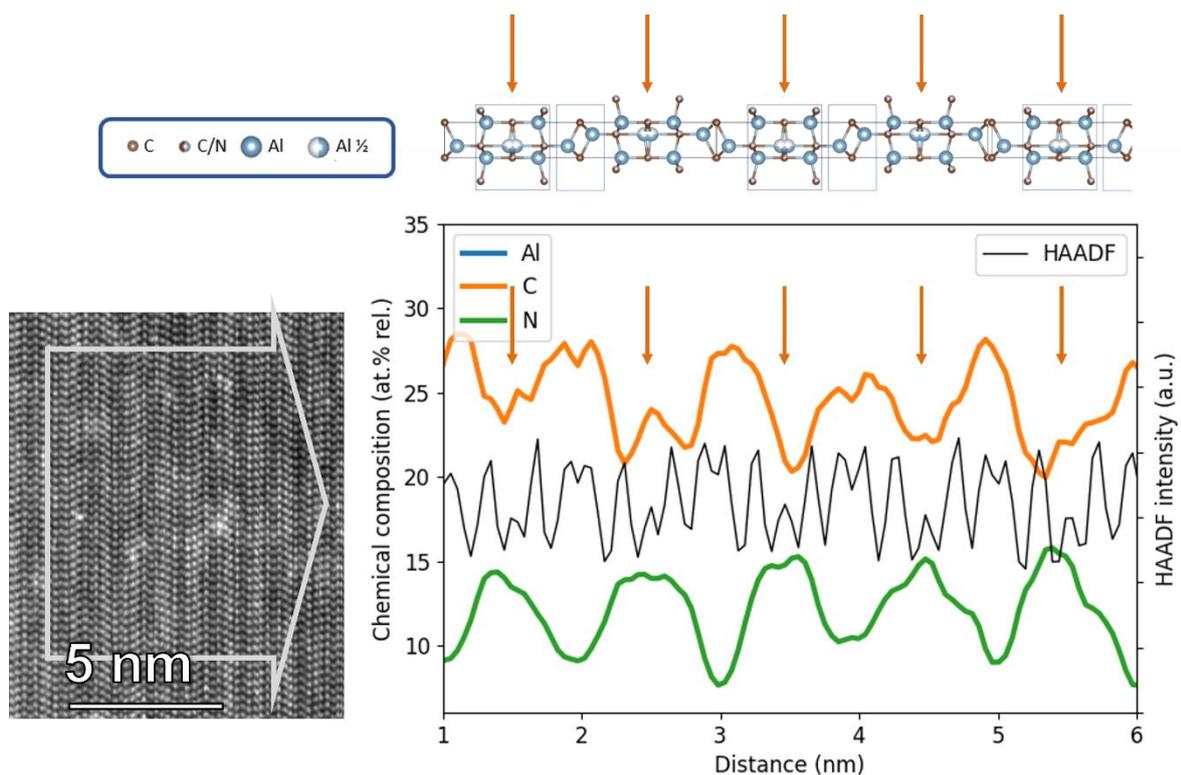

*Figure SI6.* Identical dataset as presented in **Error! Reference source not found.** but processed with a much lower binning rate (5 pixels instead of 11). The atomic model is scaled and aligned to the composition profile. When the binning is reduced, carbon signals starts to be visible in several Al(C,N) planes as indicated by the orange arrows. Nonetheless this signal is not totally consistent due to the higher signal to noise ratio.